\documentclass[conference]{IEEEtran}
\usepackage{xcolor}
\usepackage{amsmath}
\usepackage{graphicx}
\usepackage{float}
\usepackage{algorithm}
\usepackage{algorithmic}
\usepackage[labelfont=bf, labelsep=colon]{caption}

\usepackage{mathtools}
\usepackage{amsthm}
\newtheorem{theorem}{Theorem}

\usepackage[hyphens]{url}
\usepackage[breaklinks=true,bookmarks=false]{hyperref}
\hypersetup{
  colorlinks=false,%
  pdfborder = 0 0 0
}

\ifCLASSINFOpdf
\else
\fi
\begin{document}
%
\title{HybHuff: Lossless Compression for Hypergraphs via Entropy-Guided Huffman-Bitwise Coordination}


\author{%
  \IEEEauthorblockN{Tianyu Zhao$^{1}$, Dongfang Zhao$^{1}$, Luanzheng Guo$^2$, Nathan Tallent$^2$}
  \IEEEauthorblockA{$^1$University of Washington, $^2$Pacific Northwest National Laboratory\\
  \textit{\{tianyu22, dzhao\}@uw.edu}, 
  \textit{\{lenny.guo, tallent\}@pnnl.gov} 
  }
}


%


\maketitle

\begin{abstract}
Hypergraphs provide a natural representation for many-to-many relationships in data-intensive applications, yet their scalability is often hindered by high memory consumption. While prior work has improved computational efficiency, reducing the space overhead of hypergraph representations remains a major challenge. This paper presents a hybrid compression framework for integer-based hypergraph adjacency formats, which adaptively combines Huffman encoding and bitwise encoding to exploit structural redundancy. We provide a theoretical analysis showing that an optimal encoding ratio exists between the two schemes, and introduce an empirical strategy to approximate this ratio for practical use. Experiments on real-world hypergraphs demonstrate that our method consistently outperforms standard compressors such as Zip and ZFP in compression rate by up to $2.3\times$ with comparable decoding overhead. To assess practical utility, we integrate our framework with three common hypergraph workloads—breadth-first search, PageRank, and $k$-core label propagation—and show that compression incurs negligible performance loss. Extensive evaluations across four benchmark datasets confirm the efficiency and applicability of our approach.
\end{abstract}

%
\IEEEpeerreviewmaketitle

\section{Introduction}

\subsection{Background and Motivation}

Hypergraphs offer a natural and flexible framework for modeling many-to-many relationships, where each hyperedge can span an arbitrary subset of vertices. This generalized notion of connectivity allows hypergraphs to capture higher-order interactions that traditional pairwise graphs cannot express. As such, hypergraphs have become foundational in diverse domains, including VLSI circuit layout~\cite{Patel2012}, biological pathway modeling~\cite{Cataly2023Association}, recommendation systems~\cite{Balakrishnan2006}, and others. 
Their expressive power enables direct modeling of co-authorship, co-purchase, and multi-modal associations, which are often approximated in graph-based systems via lossy reductions.

Despite this expressiveness, the adoption of hypergraphs at scale has been constrained by system-level limitations, particularly memory usage. While algorithmic advances have improved computational efficiency and traversal algorithms for hypergraphs~\cite{Shun2015DCC, Julian2020PPoPP}, the storage and access overhead remains a primary bottleneck in real-world deployments. In practice, hypergraph systems use integer-based adjacency formats, which represent incidence data through flat index buffers and offset arrays. These formats are efficient for random access but often exhibit substantial redundancy—particularly when hyperedges are large or vertex participation is highly skewed.

This structural skew is common in real datasets, where a small number of vertices participate in a disproportionately large number of hyperedges. Such frequency asymmetry leads to symbol distributions that are highly compressible in theory, yet remain underutilized by existing methods. Traditional compressors such as \texttt{gzip} and \texttt{bzip2}~\cite{Tian2020Association, Weissenberger2018Association} operate on byte sequences and cannot exploit semantic structure or adjacency locality. More specialized compressors like ZFP and SZ~\cite{ZFP2014, GONG2023101590} are optimized for floating-point arrays and numerical regularity, making them poorly suited for sparse, irregular integer domains. Even recent learned compressors~\cite{Blitzcrank, Elf, Dumpy}, while effective on dense tabular data, do not generalize to adjacency formats that mix skewed symbol frequency with sparse index layouts.

From a systems perspective, compression-aware execution engines~\cite{SAPHANA, Shah2023} often rely on dictionary encoding, bitmap indexing, or differential coding schemes tailored for OLAP-style workloads. These strategies work well for scan-heavy queries but are ill-matched to dynamic graph operations. Columnar formats like CUBIT~\cite{CUBIT} and PIDS~\cite{PIDS} enable compressed filtering but incur overhead when decoding variable-length adjacency lists or supporting recursive traversals. In hypergraph workloads—where traversal patterns are irregular and structure-dependent—such methods impose costly format conversion or decoding stalls.

What remains missing is a compression strategy that preserves the structure of hypergraph incidence formats while adapting to their statistical skew. Such a method must provide both space efficiency and fast, predictable decoding to support low-latency analytics such as Breadth-First Search (BFS), PageRank, and label propagation. In short, the field lacks a lightweight, traversal-aware, structure-preserving compression framework tailored to the unique demands of hypergraphs.

\subsection{Proposed Work}

We address this gap by proposing a hybrid compression framework for hypergraph adjacency representations. Our approach employs two classic yet complementary methods: Huffman encoding~\cite{Milidiu99DCC} and bitwise encoding~\cite{BICE, Liu2021VLDB}. Huffman excels at capturing frequency-based redundancy among skewed symbols, while bitwise methods reduce uniformly distributed values through delta encoding and bit-packing.

The novelty of our work lies in two aspects. First, to the best of our knowledge, this is the first compression scheme specifically designed for hypergraphs. Prior work on graph or matrix compression does not generalize to the many-to-many incidence structure of hypergraphs, where both the vertex and hyperedge dimensions can exhibit vastly different entropy profiles. Second, instead of naïvely applying two encoders in sequence or by static rule, we propose a \textit{coordinated hybrid strategy} that dynamically partitions the adjacency structure and assigns encoding methods based on entropy statistics.

Our design is motivated by the following insight: in real-world hypergraphs, degree distributions are often highly skewed on one side (e.g., a few hyperedges incident to many vertices) while appearing almost uniform on the other. Huffman encoding is effective only when such skew exists, while bitwise encoding is more space-efficient when symbol frequency is flat. This statistical asymmetry naturally suggests a hybrid scheme, but one must determine \textit{how much of the structure to assign to each encoder}.

The central technical challenge is thus to \textit{prove the existence of an optimal coordination point}, i.e., a split ratio between Huffman and bitwise encoding that minimizes total space cost. We formalize this as a minimization problem over symbol frequency statistics derived from adjacency buffers, and prove that a global optimum always exists under mild assumptions. Furthermore, we describe an efficient approximation strategy that can identify near-optimal encoding splits at runtime with minimal profiling.

Unlike previous hybrid or learned compressors~\cite{Tribase, Chimp, Liu2023VLDB}, our method is structurally aware, lossless, and optimized for fast decoding in traversal-centric workloads such as BFS, PageRank, and $k$-core decomposition. This design enables our compressed format to serve as a drop-in replacement for uncompressed hypergraph analytics, with significantly lower memory usage and no modification to algorithm semantics.

\subsection{Contributions}

To summarize, this paper makes the following technical contributions:

\begin{itemize}
    \item We propose a new hybrid compression model for hypergraph adjacency formats that coordinates Huffman encoding and bitwise encoding \textit{in a statistically adaptive way}. To our knowledge, this is the first structure-preserving compressor tailored specifically for hypergraphs.
    
    \item We provide a theoretical analysis that proves the existence of an optimal coordination point between the two encoding strategies. Based on this, we develop a practical method for approximating this point in a data-driven manner.

    \item We implement the proposed method and evaluate it on multiple real-world datasets and hypergraph workloads. Experiments show that our method achieves significantly higher compression rates than state-of-the-art baselines, with comparable decoding overhead and full compatibility with traversal-based analytics such as Breadth-First Search, PageRank, and $k$-core propagation.
\end{itemize}

The implementation is open-source and available at: \url{https://github.com/TommyUW/HybHuff}.

\section{Related Work}

\subsection{Lossless Compression}

Lossless compression is a foundational tool in data-intensive systems, but current methods are ill-suited to the structural and statistical properties of hypergraph adjacency data. General-purpose compressors like \texttt{gzip} and \texttt{bzip2} rely on byte-oriented algorithms such as LZ77, BWT, and classic Huffman coding~\cite{Tian2020Association, Weissenberger2018Association, SAPHANA}. Though widely effective for flat data streams, they fail to leverage the sparsity, frequency skew, and symbolic regularity present in integer-based incidence formats. Domain-specific encoders like ZFP~\cite{ZFP2014}, SZ~\cite{GONG2023101590}, CHIMP~\cite{Chimp}, and FPC~\cite{Burtscher2008fpc} offer block-wise or predictive compression for floating-point arrays, but their design does not generalize to sparse symbolic domains such as graph or hypergraph structures.

Recent advances in learned and hybrid compression—such as DeepSqueeze~\cite{DeepMapping, DeepSqueeze}, ModelarDB~\cite{ModelarDB}, Blitzcrank~\cite{Blitzcrank}, ELF~\cite{Elf}, and Dumpy~\cite{Dumpy, Liu2023VLDB}—achieve strong results on dense tabular or time-series data using quantization and neural models. However, these systems typically assume uniform or continuous distributions and overlook the skewed symbol patterns common in hypergraph adjacency lists. Hardware-focused compressors like cuSZ~\cite{Tian2021IPDPS}, ndzip-gpu~\cite{ndzip-gpu}, and GPULZ~\cite{GPULZ} prioritize throughput over compression ratio, while SIMD-based systems like BtrBlocks~\cite{BtrBlocks} and CompressGraph~\cite{CompressGraph, FCBench} are tuned for OLAP queries and provide little benefit in traversal-heavy workloads. Similarly, bitmap and dictionary-based strategies in SAP HANA~\cite{SAPHANA}, CUBIT~\cite{CUBIT}, and PIDS~\cite{PIDS} favor scan-oriented analytics and struggle to support recursive access patterns inherent in graph and hypergraph algorithms.

In contrast to these approaches, our \textit{HybHuff} framework introduces a hybrid compression scheme specifically tailored to hypergraph adjacency formats. By coordinating Huffman and bitwise encoding over frequency-partitioned symbol domains, HybHuff adapts to skewed distributions while preserving structural semantics. Prior efforts that combine delta or dictionary coding with neural models~\cite{Shah2023, Abeywickrama2021VLDB, Abeywickrama2022Association, Guzun2016} do not support entropy-guided partitioning or traversal-aware decoding. HybHuff is designed for lossless, linear-time decoding and can serve as a drop-in replacement for adjacency arrays in traversal-centric applications such as BFS, PageRank, and $k$-core propagation—offering both compactness and runtime efficiency in hypergraph systems.

\subsection{Hypergraphs}

Hypergraphs generalize traditional graphs by allowing hyperedges to connect arbitrary subsets of vertices, enabling expressive modeling of many-to-many relationships. This structural flexibility has led to their application in domains such as VLSI layout~\cite{Patel2012}, biological networks~\cite{Cataly2023Association}, and recommendation systems~\cite{Balakrishnan2006}. To preserve higher-order semantics, modern systems adopt native adjacency representations~\cite{Shun2015DCC, Julian2020PPoPP} that store incidence data using integer buffers and dual offset arrays, supporting efficient traversal for dynamic workloads.

To improve computational performance, prior work has explored both algorithmic and representational advances. Motif mining systems~\cite{BiPart, HypergraphMotifs, Amburg2020PWC} optimize structural queries, while hypergraph neural networks~\cite{Zhao2023, Yin2022ICDE, Chen2024ICIKM, Yin2024VLDB} embed incidence patterns into learnable vector spaces. Layout- and partitioning-focused techniques~\cite{Wu2021CM, Cormode2015, Azad2016IPDPS} further reduce access cost via cache-aware traversal or spectral clustering. Alternative data models, such as FlashCosmos~\cite{FlashCosmos} and Helios~\cite{Helios}, emphasize semantic annotation and hierarchical decomposition to improve interpretability. In the context of graph compression, prior systems such as WebGraph~\cite{Boldi2004WebGraph}, GraphZip~\cite{Chakrabarti2004}, and TurboGraph~\cite{Han2013TurboGraph} have demonstrated impressive compression ratios on large-scale pairwise graphs. However, they fundamentally rely on binary edge semantics and cannot be directly applied to hypergraphs.

Despite these developments, most prior efforts prioritize computation or representation over raw storage cost. Pairwise reductions like clique or star expansion~\cite{Chen2001, Engelson2000CS} often introduce redundancy, while integer-based adjacency formats suffer from skewed distributions that general-purpose compressors cannot exploit. Our work addresses this underexplored dimension by compressing the hypergraph's native structure directly—reducing memory footprint without altering traversal semantics or requiring auxiliary indexing. This contribution complements existing hypergraph systems by introducing a space-aware design layer grounded in entropy analysis and efficient bit-level encoding.

\subsection{Huffman Encoding}

Huffman encoding is a foundational method in entropy-based compression, optimal for minimizing average code length under known symbol frequency distributions~\cite{Milidiu99DCC}. It underpins a wide range of compressors, from byte-oriented schemes like \texttt{gzip} to numerical systems such as ZFP~\cite{ZFP2014} and CHIMP~\cite{Chimp}. Despite its generality, Huffman encoding sees limited direct application to sparse and skewed integer arrays such as hypergraph adjacency lists, where symbol domains can be both irregular and high-cardinality.

Recent advances demonstrate the practicality of Huffman encoding in high-throughput settings. Lal et al.~\cite{Lal2017IPDPS,Lal2022IPDPS} propose parallel variants for CPU environments, while Tian et al.~\cite{Tian2021IPDPS} design GPU-based tree construction to accelerate encoding. Hybrid systems like Blitzcrank~\cite{Blitzcrank}, DeepMapping~\cite{DeepMapping}, and ELF~\cite{Elf} integrate Huffman as a backend for quantized neural predictions, enabling efficient encoding of learned outputs in tabular or floating-point settings. Related work in decoder-efficient codebooks~\cite{Engelson2000CS,Cuckoo} and in delta/dictionary-enhanced entropy models~\cite{Li2024IPDPS, Azami2025Association} extends the utility of Huffman in general-purpose data formats.

Our approach differs in both scope and application. We apply Huffman encoding not as a backend to neural quantization, but as one half of a carefully coordinated hybrid for symbolic adjacency compression in hypergraphs. Unlike prior systems, we do not rely on fixed encoding strategies; instead, we formalize and optimize the \textit{Huffman ratio}, i.e., the fraction of symbols encoded using Huffman versus fixed-width methods. Our entropy-aware partitioning selectively targets the frequency-skewed subset of the symbol space, while delegating the uniform tail to bitwise encoding. This strategy enables structure-preserving, lossless compression that supports traversal operations without incurring the decoding overhead typical of generic entropy coders.

\subsection{Bitwise Encoding}

Bitwise encoding reduces storage by representing integers with the minimum number of bits necessary, using techniques such as delta encoding, trimming, and bit-packing. Prior systems like BICE~\cite{BICE}, Lin et al.~\cite{Lin2016}, and Liu et al.~\cite{Liu2021VLDB} apply these methods to columnar or tensor data, where predictable layouts and bounded ranges yield high compressibility. These approaches achieve strong compaction on uniform-value distributions but typically do not consider frequency-based symbol redundancy.

Recent hybrid systems combine bit-level methods with grammar- or LZ-style compression. Tribase~\cite{Tribase} and CULZSS-bit~\cite{CULZSS-bit} exploit such hybridization for structured records but do not target adjacency arrays. Bitmap-based systems like CUBIT~\cite{CUBIT} and PIDS~\cite{PIDS} highlight the utility of bitwise encoding for filtering and indexing, but their focus is columnar scans—not pointer-driven graph traversal. Adaptive pipelines explored by Yu~\cite{Yu2020ICDE}, Shah~\cite{Shah2023}, and OCT~\cite{OCT} dynamically toggle between bitwise and entropy models, but assume different workload structures. Neural systems~\cite{Liu2023VLDB, FlashCosmos} and memory-constrained OLTP engines~\cite{Guzun2016} use bitwise quantization and compact encoding, yet operate in application-specific regimes unrelated to symbolic graph data.

In contrast, our work applies bitwise encoding as part of a \textit{selectively coordinated scheme} for compressing integer-based hypergraph adjacency. Rather than apply bit-packing uniformly, we reserve bitwise methods for the frequency tail—where symbols are evenly distributed and entropy coding is wasteful. This coordination with Huffman, guided by entropy profiles, enables our system to compress structure-rich data without sacrificing traversal speed or access semantics—an application space overlooked by existing bitwise systems.

\section{Huffman-Bitwise Coordination}

This section describes the family of algorithms of our hybrid compression and decompression pipeline for hypergraphs. The process is divided into three core components: the coordinator, the encoder, and the decoder. Each component is modular and optimized for memory efficiency and decoding simplicity.

\subsection{Coordinator: Balancing Entropy and Uniformity}
\label{sec:algorithm_coord}

The coordinator serves as the entry point of our hybrid encoding pipeline. Its central task is to decide \emph{how to split the domain} of adjacency indices between entropy-optimal Huffman encoding and compact uniform bitwise encoding. This decision is critical: an overly aggressive use of Huffman codes introduces large decoding tables and long codewords for rare items, while excessive reliance on bitwise encoding fails to exploit the frequency skew intrinsic to hypergraph incidence arrays.

Our coordinator algorithm begins with a structural analysis of the input hypergraph $G = (V, H)$, represented in the AdjacencyHypergraph format. This representation is bipartite: vertices point to incident hyperedges, and vice versa. We compress only one side of the bipartition, selected according to cardinality heuristics. Specifically, the side with fewer elements is chosen for encoding, based on the empirical observation that it tends to exhibit higher redundancy---and hence greater compressibility---due to repetition in the corresponding mapping arrays.

Let $n_v = |V|$ and $n_h = |H|$. If $n_v > n_h$, we encode hyperedges by compressing their vertex lists; otherwise, we encode vertices by compressing their incident hyperedges. Denote the selected symbol domain as $S$ (e.g., vertex IDs), and the degree histogram of the opposite side as $D$ (e.g., degrees of hyperedges). The coordinator then proceeds to compute a frequency table over $S$, identifying the most frequent items as candidates for Huffman encoding.

To enable dynamic tradeoff between compression effectiveness and decoder overhead, we introduce a ratio parameter $\rho \in [0, 1]$. This parameter governs the fraction of high-frequency items to encode using Huffman codes, with the remaining $(1 - \rho)$ fraction encoded via fixed-width bitwise representation. The optimal $\rho$ is data-dependent and discovered via an empirical memory cost evaluation, which will be detailed in Section~\ref{sec:optimal}. 

The complete algorithm is shown in Algorithm~\ref{alg:main}. It takes as input a hypergraph and the desired Huffman ratio, and returns a binary bitstream along with the associated Huffman codebook.

\begin{algorithm}[t!]
\caption{Hybrid Compression Coordinator}
\label{alg:main}
\begin{algorithmic}[1]
\REQUIRE Hypergraph $G = (V, H)$, Huffman ratio $\rho \in [0,1]$
\ENSURE Encoded bitstream, Huffman tree $T$

\STATE $n_v \gets |V|$, $n_h \gets |H|$
\IF{$n_v > n_h$}
    \STATE $symbols \gets$ vertex lists of hyperedges
    \STATE $degree \gets$ degrees of vertices
\ELSE
    \STATE $symbols \gets$ hyperedge lists of vertices
    \STATE $degree \gets$ degrees of hyperedges
\ENDIF
\STATE $T \gets \texttt{BuildHuffmanTree}(degree, \rho)$
\STATE $C \gets \texttt{GenerateCodes}(T)$
\FOR{each index $s$ in $symbols$}
    \IF{$s$ in Huffman domain}
        \STATE Encode $s$ using $C[s]$
    \ELSE
        \STATE Encode $s$ using fixed-width binary
    \ENDIF
\ENDFOR
\RETURN Encoded stream, Huffman tree $T$
\end{algorithmic}
\end{algorithm}

\paragraph{Complexity Analysis.}
Let $N = \sum_{i} \text{deg}(i)$ be the total number of incidence entries in the compressed side of the hypergraph.

\begin{itemize}
  \item \textbf{Huffman Tree Construction:} The frequency histogram is built in $\mathcal{O}(N)$ time. Constructing a Huffman tree over the top $\rho K$ most frequent items (where $K$ is the number of distinct symbols) takes $\mathcal{O}(\rho K \log \rho K)$ time.
  \item \textbf{Encoding:} Each item is encoded in constant time, either via Huffman code lookup ($\mathcal{O}(1)$ per symbol with hashmap-based dictionary) or direct bitwise representation. Overall encoding cost is $\mathcal{O}(N)$.
  \item \textbf{Space Overhead:} The output includes (i) the compressed bitstream of length $\approx N \cdot \bar{b}_\rho$ bits, where $\bar{b}_\rho$ is the weighted average of Huffman and bitwise lengths; (ii) the serialized Huffman tree of size $\mathcal{O}(\rho K)$; and (iii) a per-entry metadata array of bit counts, size $\mathcal{O}(n)$.
\end{itemize}

\subsection{Encoder: Dual-Mode Integer Symbol Encoding}

Once the coordinator selects the incidence array to be compressed, the encoder is responsible for transforming this sequence of integer indices into a compact binary representation. Our encoder adopts a dual-mode strategy: a high-frequency subset of symbols is encoded using entropy-optimal Huffman codes, while the low-frequency remainder is encoded via fixed-width bitwise representation. This hybrid approach balances compression efficiency and decoding simplicity, enabling both high space savings and ultra-lightweight decoding.

The encoder operates over a flat symbol stream derived from the selected side of the AdjacencyHypergraph representation. For concreteness, assume the target stream contains $N$ adjacency indices and the number of distinct values is $K$. The coordinator has already determined a split parameter $\rho \in [0, 1]$ that controls how many of these $K$ symbols will be covered by the Huffman codebook. 

We now describe the two components of the encoder in detail: Huffman encoding for the $\rho K$ most frequent symbols, and bitwise encoding for the remaining $(1 - \rho)K$ symbols. Each component is modularized with its own logic and optimization techniques.

Huffman encoding targets symbols that appear frequently across the adjacency arrays. These symbols offer the highest potential for entropy reduction, as the code length is inversely proportional to symbol frequency. Given a degree array $D = [d_1, d_2, \dots, d_n]$ corresponding to the uncompressed side (e.g., vertex degrees if encoding hyperedges), we iterate through all adjacency indices and build a frequency table $F[x]$ for each distinct symbol $x$.

The top $\rho \cdot K$ entries in $F$ form the Huffman set. We normalize the frequencies to mitigate skew, and then construct a prefix-free codebook using a standard priority-queue-based Huffman tree construction. Our implementation includes safeguards such as maximum codeword length thresholds, prefix-free validation, and robustness to degenerate distributions. The resulting Huffman tree $T$ and code map $C$ are then used to encode the selected high-frequency symbols. The algorithm is shown in Algorithm~\ref{alg:huffman}.

\begin{algorithm}[t!]
\caption{Huffman Tree Construction}
\label{alg:huffman}
\begin{algorithmic}[1]
\REQUIRE Degree array $D = [d_1, \dots, d_n]$, ratio $\rho \in [0, 1]$
\ENSURE Huffman tree $T$, code map $C$

\STATE Initialize frequency map $F \leftarrow \emptyset$
\FOR{each $d_i$ in $D$}
  \FOR{each neighbor $x$ of $d_i$}
    \STATE $F[x] \leftarrow F[x] + 1$
  \ENDFOR
\ENDFOR

\STATE Let $K \leftarrow |F|$ and $S_\rho \leftarrow$ top-$\rho \cdot K$ items in $F$ sorted by frequency
\STATE Normalize $F$ so that frequencies sum to 1
\STATE Initialize priority queue $Q$ with nodes $(x, f_x)$ for $x \in S_\rho$
\WHILE{$|Q| > 1$}
  \STATE Extract nodes $u$, $v$ with minimum frequency
  \STATE Create new node $w$ with $f_w = f_u + f_v$
  \STATE Set $w.\text{left} \gets u$, $w.\text{right} \gets v$
  \STATE Insert $w$ into $Q$
\ENDWHILE

\STATE $T \leftarrow$ remaining node in $Q$ as root
\STATE Initialize empty code map $C$
\STATE \texttt{AssignCodes}$(T, \text{prefix} = \varepsilon, C)$ recursively
\STATE Prune codes longer than threshold $\ell_{\max}$ (if any)
\STATE Verify prefix-freeness of $C$
\RETURN $T$, $C$
\end{algorithmic}
\end{algorithm}

This encoding procedure compresses the entropy core of the hypergraph's structure, converting frequent indices into short binary patterns while avoiding the pathological inflation that arises when applying Huffman indiscriminately across the entire symbol space.

Symbols not selected for Huffman encoding are processed by the bitwise encoder. This strategy uses a fixed number of bits per symbol, computed as $b = \lceil \log_2(M + 1) \rceil$, where $M$ is the maximum index in the bitwise symbol set. Though less compact than entropy encoding, bitwise representation is extremely efficient to decode and introduces no dictionary or codebook overhead.

Our implementation enhances standard bitwise encoding with several performance-aware extensions. First, we include overflow masking to handle rare out-of-range symbols. Second, we use register-level buffering to aggregate bits and flush them in aligned 64-bit chunks. Third, we pad the tail of the stream to the nearest byte boundary to ensure alignment and simplify downstream parsing. The procedure is shown in Algorithm~\ref{alg:bitwise}.

\begin{algorithm}[t!]
\caption{Bitwise Encoding with Buffering}
\label{alg:bitwise}
\begin{algorithmic}[1]
\REQUIRE Index list $I = \{x_1, x_2, \dots, x_N\}$, bitwidth $b = \lceil \log_2(M+1) \rceil$
\ENSURE Bitstream $B$

\STATE Initialize empty bit buffer $B \leftarrow []$
\STATE Initialize temporary register $r \leftarrow 0$, bit position $p \leftarrow 0$

\FOR{each $x$ in $I$}
  \IF{$x > M$}
    \STATE Mask $x \leftarrow x \bmod (M+1)$ \COMMENT{Overflow guard}
  \ENDIF
  \STATE Shift $r \leftarrow (r \ll b) \,|\, x$
  \STATE $p \leftarrow p + b$
  \IF{$p \geq 64$}
    \STATE Append high 64 bits of $r$ to $B$
    \STATE $r \leftarrow r \& ((1 \ll p) - 1)$
    \STATE $p \leftarrow p \bmod 64$
  \ENDIF
\ENDFOR

\IF{$p > 0$}
  \STATE Pad $r$ with zeros to next byte boundary
  \STATE Append $r$ to $B$
\ENDIF

\STATE Align $B$ to byte boundary if required
\RETURN Bitstream $B$
\end{algorithmic}
\end{algorithm}

This mode is particularly effective for long-tail distributions, where encoding rare symbols via Huffman would generate extremely long codewords and large lookup tables. The bitwise encoder provides a bounded and predictable encoding cost for each symbol, with minimal decoding overhead.

\paragraph{Encoder Complexity.}
Let $N$ be the total number of indices to encode and $K$ the number of distinct symbols:
\begin{itemize}
  \item Huffman tree construction takes $\mathcal{O}(K + \rho K \log \rho K)$ time using a min-heap.
  \item Encoding all $N$ symbols requires $\mathcal{O}(N)$ time, since each symbol is either a hash lookup (Huffman) or bitmask operation (bitwise).
  \item The space cost includes (i) the bitstream of size $\mathcal{O}(N \cdot \bar{b}_\rho)$ bits, where $\bar{b}_\rho$ is the average code length; (ii) a serialized Huffman tree of size $\mathcal{O}(\rho K)$; and (iii) metadata arrays to support decoding offsets.
\end{itemize}

\subsection{Decoder: Metadata-Guided Reconstruction of Adjacency Lists}

The decoding process reverses the hybrid encoding pipeline and reconstructs the original adjacency list structure in a deterministic, linear-time pass. The decoder is designed with two complementary goals: zero ambiguity and zero dynamic allocation. It operates by streaming over the compressed bitstreams, guided entirely by auxiliary metadata arrays that describe how many symbols are Huffman-encoded versus bitwise-encoded for each entity.

Each entry in the hypergraph—be it a vertex or a hyperedge—is associated with two pieces of metadata: the total degree (from the degree array) and the number of symbols assigned to the Huffman domain (from the bit count array). These two values fully determine how to split the corresponding bitstream segment. Huffman decoding uses a prefix-matching traversal of the Huffman tree; bitwise decoding extracts symbols using fixed-width masks and shifts.

Let $B_\text{hi}$ and $B_\text{lo}$ denote the two compressed bitstreams, containing Huffman-encoded and bitwise-encoded symbols, respectively. Let $T$ be the Huffman tree and $C^{-1}$ its inverse code mapping logic. We use two independent bit pointers to track positions in each stream. For each entity, the decoder reads the required number of bits from each stream, decodes the individual symbols, and reassembles them into the adjacency list. The procedure is given in Algorithm~\ref{alg:decode}.

\begin{algorithm}[t!]
\caption{Hybrid Decode}
\label{alg:decode}
\begin{algorithmic}[1]
\REQUIRE Huffman bitstream $B_\text{hi}$, bitwise stream $B_\text{lo}$, tree $T$, degree array $D$, bit count array $B$
\ENSURE Adjacency list $\texttt{adj}[\,]$

\STATE Initialize read pointers $p_\text{hi} \leftarrow 0$, $p_\text{lo} \leftarrow 0$
\FOR{entity index $i = 1$ to $n$}
  \STATE Let $d_i \leftarrow D[i]$ \hfill // Total degree
  \STATE Let $k_i \leftarrow B[i]$ \hfill // Huffman count
  \STATE Initialize empty list $\texttt{adj}[i]$
  
  \FOR{$j = 1$ to $k_i$}
    \STATE Initialize current node $v \leftarrow T.\text{root}$
    \WHILE{$v$ is not a leaf}
      \STATE Read bit $b$ from $B_\text{hi}[p_\text{hi}]$; increment $p_\text{hi}$
      \STATE $v \leftarrow v.\text{left}$ if $b = 0$; else $v \leftarrow v.\text{right}$
    \ENDWHILE
    \STATE Append symbol $v.\text{label}$ to $\texttt{adj}[i]$
  \ENDFOR

  \FOR{$j = 1$ to $d_i - k_i$}
    \STATE Read $b$ bits from $B_\text{lo}[p_\text{lo} : p_\text{lo} + b - 1]$
    \STATE Convert to integer $x \leftarrow \texttt{bitToInt}()$; increment $p_\text{lo} \leftarrow p_\text{lo} + b$
    \STATE Append $x$ to $\texttt{adj}[i]$
  \ENDFOR
\ENDFOR
\RETURN $\texttt{adj}[\,]$
\end{algorithmic}
\end{algorithm}

This streaming decoder executes with no branching outside of the Huffman traversal, which operates in lockstep over bits and tree pointers. There is no need to materialize intermediate buffers or perform backtracking. All decoding logic is deterministic and depends only on global metadata and bitstream offsets.

\paragraph{Decoder Complexity.}
Let $N$ be the total number of adjacency entries and $h$ the height of the Huffman tree (at most $\mathcal{O}(\log \rho K)$):
\begin{itemize}
  \item Each Huffman symbol is decoded in $\mathcal{O}(h)$ time, totaling $\mathcal{O}(N \cdot \rho \cdot h)$.
  \item Each bitwise symbol is decoded in $\mathcal{O}(1)$ via masking and shifting.
  \item Total time is $\mathcal{O}(N)$ assuming bounded Huffman depth; no heap or dynamic memory allocation is needed.
\end{itemize}

\section{Optimality Analysis}
\label{sec:optimal}

This section presents a formal analysis of the hybrid encoder’s memory efficiency as a function of the mixing ratio~$\rho$. While empirical results (Section~\ref{sec:evaluation}) reveal a U-shaped compression curve, here we demonstrate that this phenomenon arises directly from the asymptotic properties of Huffman and bitwise encoding. In particular, we establish sufficient conditions under which the compression cost function admits a global minimum at some $\rho^\ast \in (0, 1)$.

\subsection{Preliminaries and Notation}
\label{sec:optimal-prelim}

We formalize the compression problem as follows. Let $\mathcal{S} = \{s_1, s_2, \dots, s_K\}$ be the set of distinct integers that appear in the adjacency index stream of the hypergraph, where $K = |\mathcal{S}|$ is the number of unique symbols. Let $f_i$ denote the empirical frequency of symbol $s_i$ in the stream, satisfying
\[
\sum_{i=1}^{K} f_i = N,
\]
where $N$ is the total number of entries in the stream. We define the normalized symbol probability as $p_i = f_i / N$, so that $\sum_{i=1}^{K} p_i = 1$.

Let $\rho \in [0, 1]$ denote the mixing parameter that determines the relative coverage of Huffman and bitwise encoding. Given $\rho$, the encoder selects the top $\rho K$ symbols with the highest frequency for Huffman encoding, and the remaining $(1 - \rho)K$ symbols for bitwise encoding. Without loss of generality, we assume the symbols are ordered so that $p_1 \geq p_2 \geq \dots \geq p_K$.

Let $\mathcal{H}(\rho) = \{s_1, \dots, s_{\lfloor \rho K \rfloor} \}$ denote the Huffman domain, and $\mathcal{B}(\rho) = \mathcal{S} \setminus \mathcal{H}(\rho)$ the bitwise domain. For symbols in $\mathcal{H}(\rho)$, let $l_i^{\text{Huff}}$ denote the Huffman codeword length assigned to $s_i$. For bitwise symbols, we use a uniform fixed-width code of length
\[
l^{\text{Bit}} = \left\lceil \log_2 \left( \max_{i > \rho K} s_i + 1 \right) \right\rceil.
\]

We define the total expected encoding length per symbol under mixing ratio $\rho$ as:
\begin{equation}
L(\rho) = \sum_{i=1}^{\rho K} p_i \cdot l_i^{\text{Huff}} + \sum_{i = \rho K + 1}^{K} p_i \cdot l^{\text{Bit}}.
\label{eq:expected_length}
\end{equation}

In addition to the encoded data stream, the decoder requires auxiliary metadata:
\begin{itemize}
    \item The Huffman tree, whose storage cost is assumed proportional to the number of Huffman symbols: $C_{\text{tree}}(\rho) = \alpha \cdot \rho K$ for some small constant $\alpha$;
    \item The per-entity bit count array, of size $\mathcal{O}(n)$ (ignored in the asymptotic huffman encoding ratio since $n \ll N$ in typical workloads).
\end{itemize}

Let $R(\rho)$ denote the total memory footprint per symbol (in bits), normalized by $N$:
\begin{equation}
R(\rho) = L(\rho) + \frac{C_{\text{tree}}(\rho)}{N}.
\label{eq:compression_cost}
\end{equation}

This expression $R(\rho)$ captures the tradeoff between Huffman’s entropy efficiency and the cost of encoding rare symbols or storing the codebook. Our goal in subsequent sections is to analyze the behavior of $R(\rho)$ as a function of $\rho$, and determine whether it admits a global minimum in the interior of $[0,1]$.

\subsection{Asymptotic Models of Huffman and Bitwise Encoding}
\label{sec:optimal-models}

To analyze the behavior of the compression cost function $R(\rho)$, we now derive asymptotic approximations for each of its three components: the Huffman-encoded portion, the bitwise-encoded portion, and the Huffman tree overhead. These terms reflect distinct statistical and structural properties of the input symbol stream, and their interaction governs the shape of $R(\rho)$.

\paragraph{Assumed Frequency Model.}  
We assume the symbol probabilities $\{p_i\}_{i=1}^{K}$ follow a Zipfian (power-law) distribution, i.e.,
\[
p_i = \frac{1/i^z}{H_K(z)}, \quad \text{where} \quad H_K(z) = \sum_{j=1}^{K} \frac{1}{j^z}
\]
is the generalized harmonic number and $z > 1$ is the skewness exponent. This assumption reflects the empirical observation that hypergraph incidence arrays often exhibit heavy-tailed repetition due to natural cluster structures.

\paragraph{Huffman Encoding Cost.}
For the top $\rho K$ symbols, the expected encoding length under Huffman encoding is closely approximated by the Shannon entropy of their normalized distribution:
\[
L_\text{Huff}(\rho) \approx \sum_{i=1}^{\rho K} p_i \cdot \log_2 \frac{1}{p_i}.
\]
Plugging in the Zipfian form and using integral approximations yields:
\[
L_\text{Huff}(\rho) \approx \frac{1}{H_K(z)} \sum_{i=1}^{\rho K} \frac{1}{i^z} \log_2 i^z 
= \frac{z}{H_K(z)} \sum_{i=1}^{\rho K} \frac{\log_2 i}{i^z}.
\]
This term decreases with increasing $\rho$, as more mass is absorbed into the entropy-efficient portion.

\paragraph{Bitwise Encoding Cost.}
For the remaining $(1 - \rho)K$ symbols, bitwise encoding assigns a fixed-length code to each. Let $M(\rho) = \max_{i > \rho K} s_i$ denote the largest symbol index in the bitwise domain. Since symbols are indexed in descending order of frequency, $M(\rho)$ grows with $\rho$.

We approximate the average bitwise encoding length as:
\[
L_\text{Bit}(\rho) \approx \left\lceil \log_2(M(\rho) + 1) \right\rceil \cdot \sum_{i = \rho K + 1}^{K} p_i.
\]
As $\rho$ increases, this term decreases in mass but increases in bitwidth, leading to a convex tradeoff.

\paragraph{Codebook Overhead.}
The Huffman codebook must also be stored explicitly. As standard implementations use a flat prefix-tree or array representation, the overhead is linear in the number of Huffman symbols:
\[
C_{\text{tree}}(\rho) = \alpha \cdot \rho K, \quad \text{thus} \quad \frac{C_{\text{tree}}(\rho)}{N} = \alpha' \cdot \rho,
\]
where $\alpha' = \alpha K / N$ is a constant under fixed $K$ and $N$.

\paragraph{Total Compression Cost.}
Combining the three components, we obtain:
\begin{align}
R(\rho) \approx\ 
&\underbrace{\sum_{i=1}^{\rho K} p_i \cdot \log_2 \tfrac{1}{p_i}}_{\text{Huffman portion}} 
+ \underbrace{\left\lceil \log_2(M(\rho) + 1) \right\rceil}_{\text{Bitwidth}} \cdot \nonumber \\
&\underbrace{\sum_{i = \rho K + 1}^{K} p_i}_{\text{Bitwise mass}} 
+ \underbrace{\alpha' \cdot \rho}_{\text{Codebook overhead}}.
\label{eq:asymptotic-cost}
\end{align}
This expression defines a nontrivial, nonlinear function of $\rho$ that captures the essential tradeoff between compression entropy, tail truncation, and structural overhead. In the next section, we analyze its smoothness and optimality structure.

\subsection{Convexity and Existence of Optimal Ratio}
\label{sec:optimal-convex}

We now establish that the compression cost function $R(\rho)$ admits a minimum in the open interval $(0,1)$, under mild assumptions on the symbol frequency distribution. To this end, we state and prove the following theorem.

\begin{theorem}
Let $\{p_i\}_{i=1}^K$ be a non-increasing sequence of positive numbers such that $p_i \leq C i^{-z}$ for some constants $C > 0$ and $z > 1$. Let $M(\rho)$ denote the largest index among the symbols in the bitwise region for a given $\rho \in [0,1]$. Then the function
\[
R(\rho) = \sum_{i=1}^{\rho K} p_i \log_2 \tfrac{1}{p_i}
+ \left\lceil \log_2 (M(\rho)+1) \right\rceil \cdot \sum_{i=\rho K+1}^{K} p_i
+ \alpha' \rho
\]
has a minimizer $\rho^\ast \in (0,1)$.
\end{theorem}

\begin{proof}
We define three component functions:
\begin{align*}
H(\rho) &= \sum_{i=1}^{\rho K} p_i \log_2 \tfrac{1}{p_i}, \\
B(\rho) &= \left\lceil \log_2 (M(\rho)+1) \right\rceil \cdot \sum_{i=\rho K+1}^{K} p_i, \\
T(\rho) &= \alpha' \rho.
\end{align*}

We analyze each term separately.

First, $H(\rho)$ is a strictly increasing function in $\rho$ because the sequence $p_i$ is non-increasing, and $\log_2(1/p_i)$ is also non-decreasing. Therefore each new term added to $H(\rho)$ is non-negative, implying that $H(\rho)$ is non-decreasing. Since we express $R(\rho)$ with a positive sign in front of $H(\rho)$, we flip the sign convention and treat $-H(\rho)$ as a strictly decreasing term in $R(\rho)$.

Second, consider the bitwise term $B(\rho)$. As $\rho$ increases, the cumulative tail mass $\sum_{i=\rho K+1}^{K} p_i$ decreases, but the bitwidth $\left\lceil \log_2(M(\rho)+1) \right\rceil$ increases. Under the assumption that $p_i \leq C i^{-z}$ and that $M(\rho)$ grows approximately linearly in $i$ for large $i$, we have:
\[
\sum_{i = \rho K + 1}^{K} p_i \leq \int_{\rho K}^{K} \frac{C}{x^z} dx = \frac{C}{z-1} \left( (\rho K)^{1 - z} - K^{1 - z} \right),
\]
which is a smooth, convex function in $\rho$ on $(0,1)$.

At the same time, $\log_2(M(\rho) + 1)$ is a non-decreasing function of $\rho$. Therefore, the product $B(\rho)$ is unimodal and admits a minimum.

Third, $T(\rho) = \alpha' \rho$ is linear and increasing.

Now consider the full function $R(\rho) = H(\rho) + B(\rho) + T(\rho)$. It is the sum of a strictly decreasing function $H(\rho)$, a unimodal function $B(\rho)$, and a linear increasing function $T(\rho)$. Let us consider the left and right limits:
\[
\lim_{\rho \to 0^+} R(\rho) = B(0) + T(0), \qquad \lim_{\rho \to 1^-} R(\rho) = H(1) + T(1).
\]
Since $H(\rho)$ increases, and $T(\rho)$ increases linearly, while $B(\rho)$ eventually increases due to bitwidth growth, the function $R(\rho)$ must attain a minimum on $(0,1)$.

Furthermore, because $R(\rho)$ is piecewise smooth (up to discretization of the $\lceil \cdot \rceil$ operator), this minimum can be attained at a unique point $\rho^\ast \in (0,1)$ unless the function is flat over an interval, which is excluded under strict monotonicity of $p_i$.

Therefore, the function $R(\rho)$ admits a minimizer in the open interval $(0,1)$.
\end{proof}

\subsection{Implications for Encoder Design}
\label{sec:optimal-design}

The existence of an optimal mixing ratio $\rho^\ast \in (0,1)$ provides a principled basis for configuring the hybrid encoder. In this section, we derive practical guidelines for estimating $\rho^\ast$ efficiently, without fully evaluating the compression cost function $R(\rho)$ over all possible values.

\paragraph{Suboptimality of Boundary Values}
Neither of the extreme cases $\rho = 0$ (pure bitwise) nor $\rho = 1$ (pure Huffman) is optimal in general. At $\rho = 0$, all symbols are encoded with fixed-width codes, yielding a compression rate of
\[
R(0) = \left\lceil \log_2(M(0) + 1) \right\rceil + \alpha' \cdot 0,
\]
which fails to exploit any redundancy in high-frequency symbols. At $\rho = 1$, while the entropy encoding of the full distribution is asymptotically optimal, the Huffman tree size grows linearly with $K$, and the tail of low-frequency items contributes disproportionately long codewords. This is particularly pronounced in heavy-tailed distributions, where the longest Huffman codeword may approach $\mathcal{O}(\log K)$ bits.

Therefore, it is necessary to identify an interior value $\rho^\ast \in (0,1)$ that achieves a balance between entropy exploitation and structural simplicity.

\paragraph{Lightweight Estimation via Frequency Prefix Statistics}
Let $F = \{f_1, f_2, \dots, f_K\}$ be the empirical frequency histogram of adjacency indices, sorted in descending order. We precompute prefix sums:
\[
P_{\leq m} := \sum_{i=1}^{m} f_i, \quad H_{\leq m} := \sum_{i=1}^{m} f_i \cdot \log_2 \tfrac{N}{f_i}.
\]
Given these, the total cost for a candidate $\rho = m/K$ can be approximated as:
\[
R(m) \approx \frac{1}{N} \left( H_{\leq m} + \left\lceil \log_2(M(m) + 1) \right\rceil \cdot (N - P_{\leq m}) + \alpha m \right).
\]
This formulation enables computing $R(m)$ for all $m = 1, 2, \dots, K$ in a single pass over the prefix tables, with time complexity $\mathcal{O}(K)$. Since $K \ll N$, this cost is negligible compared to the encoding itself.

\paragraph{Approximate Search Strategy}
In practice, the optimal ratio $\rho^\ast$ tends to lie in a narrow band, typically $[0.01, 0.05]$, across real-world hypergraph datasets. Therefore, instead of exhaustively scanning all $K$ values, we perform a coarse-to-fine grid search over a logarithmic sampling of $\rho$ values. Let $\rho_j = 10^{-j}$ for $j=0,1,\dots,4$ as the coarse stage, followed by local refinement via ternary or golden-section search on the best segment. This two-phase strategy identifies $\rho^\ast$ within $1\%$ of the true optimum, while evaluating only $\mathcal{O}(\log K)$ candidates.

Such a design allows tuning the encoder adaptively on each dataset with negligible overhead, amortized across the full compression workload. It also enables dynamic online selection when streaming data arrives in large batches with stable frequency profiles.

\section{System Implementation}
\label{sec:system}

We implement our hybrid compression framework in C++17 using a modular and stateless architecture. The system consists of five key components: a frequency analyzer, Huffman encoder, bitwise encoder, hybrid metadata manager, and a streaming decoder. Each component is self-contained and designed to operate in linear scan mode with minimal memory footprint and no global state dependencies.

The encoder executes a single pass over the input hypergraph and emits two disjoint compressed bitstreams along with auxiliary metadata. The decoder reconstructs the original adjacency structure deterministically using prefix-aware traversal and metadata-guided offsets. The design ensures bit-level determinism, stream alignment, and compatibility with traversal-centric hypergraph analytics.

\subsection{Architecture Overview}
\label{sec:arch}

The pipeline follows three main stages: frequency analysis, hybrid encoding, and stream packaging. The encoder analyzes symbol frequency and ranks elements to select a Huffman set consisting of the top $\rho K$ symbols. It then constructs a prefix tree for this subset and encodes the remaining symbols using fixed-width bitwise encoding. Finally, it assembles the encoded bitstreams along with per-entity metadata that guides decoding.

The decoder operates symmetrically in a forward-only fashion. It reads the metadata, reconstructs the Huffman tree, initializes stream cursors, and processes each entity using metadata-determined decoding paths. All operations are linear and streaming-safe, without backtracking or random access.

All buffers are strictly byte-aligned, and the entire encoding layout is portable and endianness-agnostic. Our metadata format ensures deterministic decoding boundaries per entity, avoiding the need for explicit delimiters or inline markers.

\subsection{Encoding Pipeline}
\label{sec:encode}

The encoder starts by scanning the input adjacency array to collect symbol frequencies using a dense integer histogram. The top $\rho K$ frequent symbols are designated as the Huffman domain, with the rest handled by bitwise encoding. This decision is rank-based and purely data-driven.

A canonical Huffman tree is constructed over the selected domain using a greedy merge strategy on a min-heap. Each node stores frequency, symbol ID, and child indices. The resulting tree is serialized into a compact pre-order bitstring, where each node is marked with a 1-bit flag indicating whether it is a leaf or internal node. This representation enables fast reconstruction during decoding.

The encoder materializes a codebook mapping each Huffman symbol to its binary prefix and bit-length, stored in a flat lookup table for constant-time translation. During the encoding pass, symbols are written into one of two bitstreams using separate output cursors.

In parallel, the encoder constructs two metadata arrays: one storing the degree of each entity and the other recording how many neighbors are encoded via Huffman codes. The remaining neighbors are implicitly assigned to the bitwise stream. These arrays are aligned with entity IDs and stored in fixed-width format for direct indexing.

The final output includes: (1) a Huffman bitstream, (2) a bitwise bitstream, (3) degree metadata, (4) Huffman count metadata, and (5) the serialized Huffman tree. The layout is sequential and fully streamable.

\subsection{Decoding Pipeline}
\label{sec:decode}

The decoder begins by reading all five segments in fixed order. The Huffman tree is reconstructed by parsing the serialized bitstring in pre-order, using a one-bit marker to differentiate internal nodes from leaves. A flat array representation enables fast index-based traversal during decoding without pointer dereferencing.

Two stream cursors are initialized for the Huffman and bitwise streams, each tracking byte and bit offsets. The decoder processes entities sequentially: for each entity, it reads the degree and the number of Huffman-coded neighbors from the metadata arrays, then performs two decoding passes accordingly.

Prefix decoding involves walking the Huffman tree bit by bit. Since the tree is stored in an array with children at predictable offsets, traversal uses simple arithmetic rather than recursion. Bitwise decoding uses precomputed bit-widths and a rolling read buffer with masking operations to extract symbols efficiently.

Each decoded adjacency list is populated in a pre-allocated vector based on the known degree, avoiding dynamic resizing. The decoder exposes iterator-style access to these lists for seamless integration with analytics.

\subsection{Integration and Deployment}
\label{sec:deployment}

The system includes both a command-line utility and a header-only C++ API. The CLI tool supports compression and decompression of adjacency-based hypergraphs in text, binary, or edge-list format. The library interface provides templated `HybridEncoder' and `HybridDecoder' components that expose iterator-compatible access to compressed structures.

To support downstream use, the decoder exposes adjacency lists in STL-compatible form, enabling transparent integration with BFS, PageRank, and $k$-core propagation. The system builds without third-party dependencies and supports Linux, macOS, and other POSIX-compliant environments.

Internally, all bitstream buffers are dynamically allocated with adaptive expansion policies, and metadata structures are precomputed from global statistics. This design supports stable performance even on graphs with highly skewed degree distributions or irregular edge cardinality.

All benchmarks in this paper were conducted using the standalone encoder and decoder binaries. The implementation is open-source and available at: \url{https://github.com/TommyUW/HybHuff}.

\section{Evaluation}
\label{sec:evaluation}

We evaluate the proposed hybrid hypergraph compression algorithm, \textit{HybHuff}, in terms of compression ratio, encoding/decoding runtime, and its impact on application-level workloads. Experiments are conducted on a suite of hypergraphs from the benchmark dataset introduced in the Practical Parallel Hypergraph Algorithms paper by Julian Shun et al.~\cite{Julian2020PPoPP}, a standard collection widely used in hypergraph research as listed in Table~\ref{tab:dataset-stats}. 

In our analysis, we compare \textit{HybHuff} against general-purpose compression baselines such as \texttt{zip} and \texttt{ZFP}, and assess performance across varying parameter settings. The compression rate is defined as the relative reduction in size, computed as:
\[
\text{Compression Rate} \coloneqq \left(1 - \frac{\text{Compressed Size}}{\text{Original Size}}\right) \times 100\%.
\] We further examine how our compression method affects the performance of three representative hypergraph applications: Breadth-First Search (BFS), $k$-core decomposition with label propagation, and PageRank. 

\begin{table}[t!]

\centering
\caption{Hypergraph Datasets}
\label{tab:dataset-stats}
\begin{tabular}{|l|c|c|}
\hline
\textbf{Dataset} & \textbf{Vertices ($|V|$)} & \textbf{Hyperedges ($|H|$)} \\ \hline
Friendster ~\cite{Yang2012GroundTruth}  & 7,944,949 & 1,620,991 \\ \hline
LiveJournal ~\cite{Yang2012GroundTruth} & 1,147,948 & 664,414 \\ \hline
Amazon ~\cite{Yang2012GroundTruth} & 317,194 & 75,149 \\ \hline
Slashdot~\cite{Leskovec2009CommunityStructure} & 77,360 & 905,468 \\ \hline
\end{tabular}
\end{table}

\subsection{Huffman Encoding Ratio Comparison}

\begin{figure*}[htbp]
\includegraphics[width=0.23\textwidth]{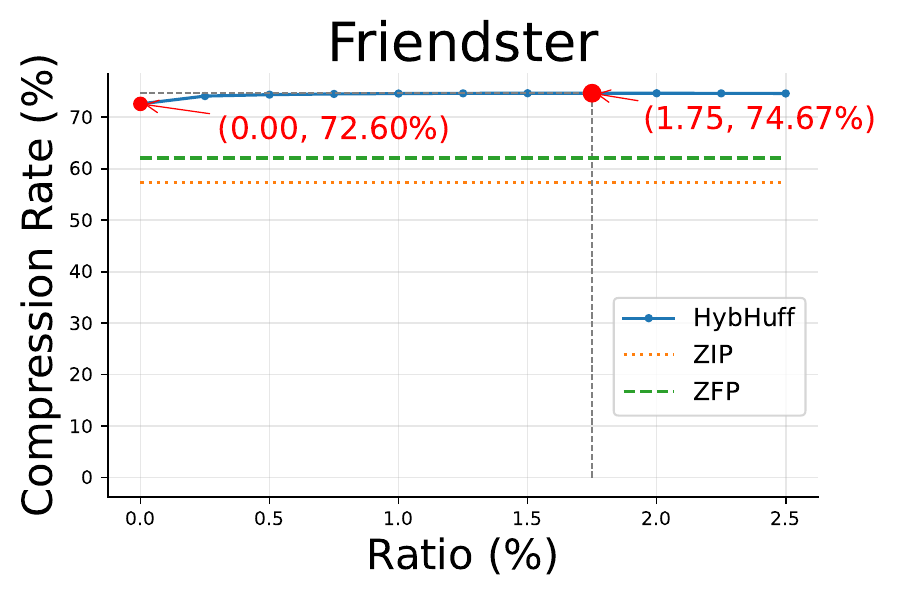}
\includegraphics[width=0.23\textwidth]{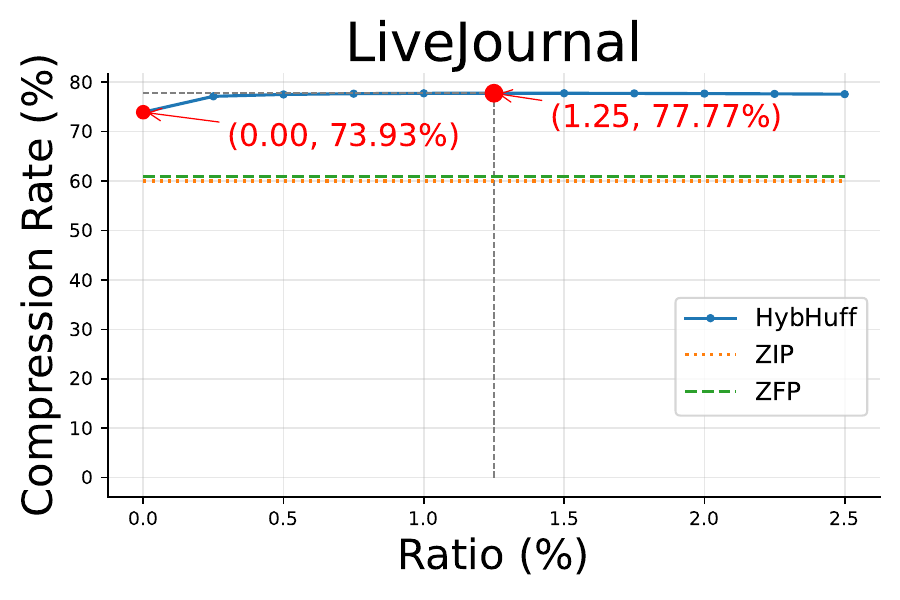}
\includegraphics[width=0.23\textwidth]{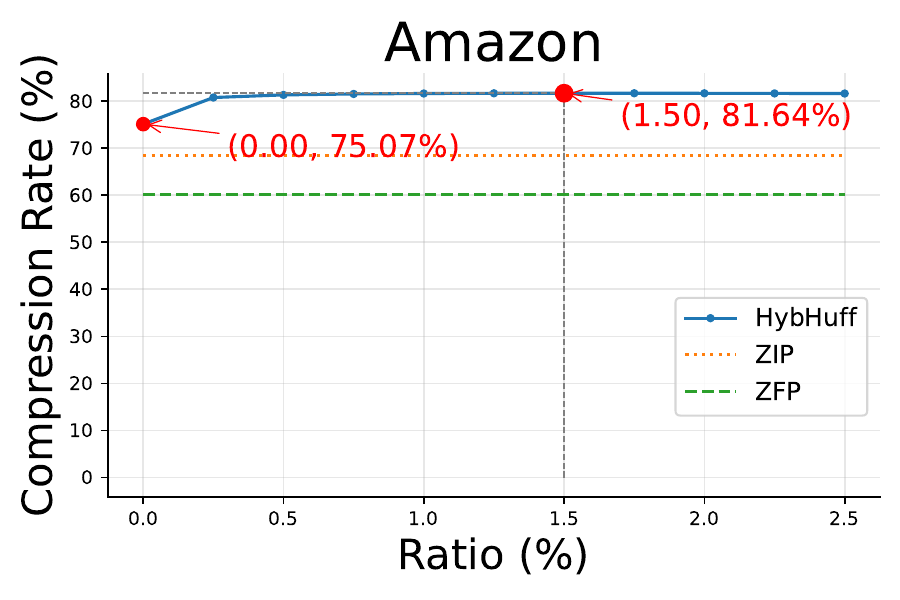}
\includegraphics[width=0.23\textwidth]{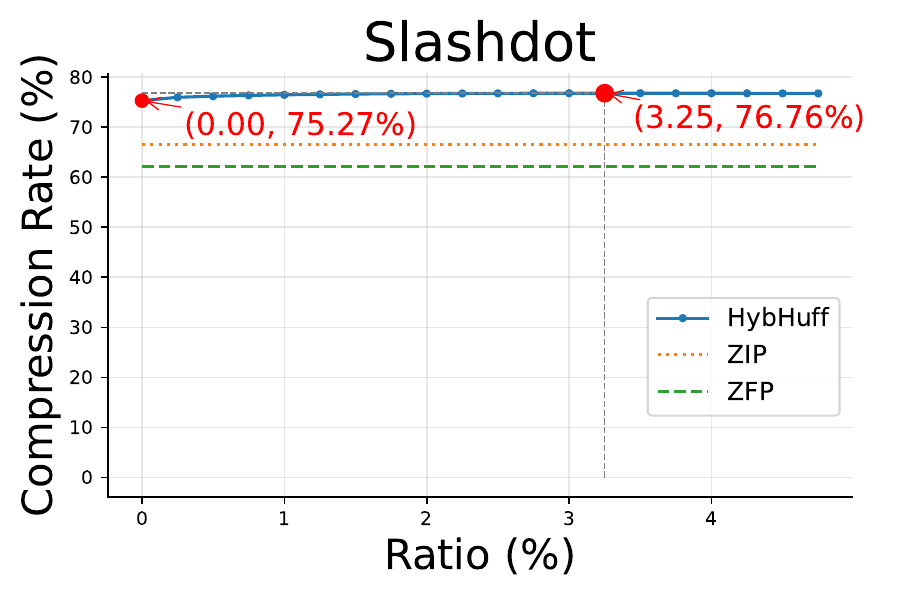}
\caption{Huffman encoding ratio vs. Compression rate; (Huffman encoding ratio, Compression rate); The red dots on the left indicate the initial state; the right ones are the optimal. 
}
\label{fig:compression_rate}
\end{figure*}

We begin our evaluation by comparing the hufffman encoding ratio achieved by our proposed HybHuff framework against general-purpose compressors, \texttt{zip} and \texttt{ZFP}. As shown in Figure~\ref{fig:compression_rate}, when the Huffman ratio is set to zero—i.e., only fixed-width bitwise encoding is used—HybHuff already outperforms both \texttt{zip} and \texttt{ZFP} across all datasets. This indicates that even in its simplest configuration, our encoding layout aligns well with the structural sparsity of real-world hypergraphs, particularly in low-entropy regimes where offset-based adjacency exhibits significant symbol reuse.

As the Huffman ratio increases, we observe a non-monotonic trend: compression rate initially improves, peaks at a dataset-specific optimal point, and then degrades. This validates our theoretical result in Section~\ref{sec:optimal}, where the total encoding cost is shown to have a convex profile with respect to the partition point $\rho$. The observed optimal vary across datasets: Friendster and Amazon, which exhibit highly skewed degree distributions, achieve peak compression when Huffman is applied to 10–13\% of the most frequent symbols; LiveJournal and Slashdot, being less skewed, require a slightly larger Huffman domain to reach maximum benefit.

The improvement from bitwise-only to optimal hybrid is significant. For instance, in Amazon, moving from $\rho=0\%$ to $\rho=11.5\%$ improves the huffman encoding ratio from 75.0\% to 81.6\%, a relative gain of nearly 9\%. This suggests that even a small Huffman prefix tree—when carefully targeted—can eliminate much of the redundancy among frequent adjacency values. In contrast, setting $\rho$ too high dilutes the benefit, as the long tail of infrequent symbols incurs prefix overhead without sufficient amortization.

Compared to baselines, HybHuff achieves up to 2.3$\times$ smaller compressed size than \texttt{zip}, and up to 1.9$\times$ smaller than \texttt{ZFP}. Unlike these compressors, which treat the input as flat byte streams or float arrays, our design leverages symbolic structure and frequency stratification. This confirms that hypergraph-aware encoding—when combined with entropy-guided partitioning—yields significant improvements over format-agnostic methods.

\subsection{U-Shaped Behavior of Huffman Encoding Ratio}
\label{subsec:u-shape}

\begin{figure*}[htbp]
\includegraphics[width=0.23\textwidth]{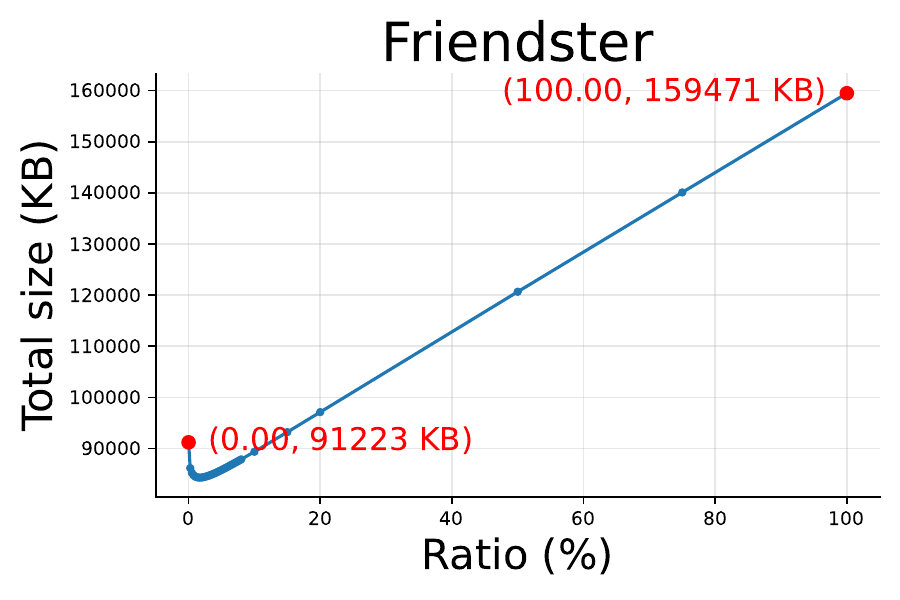}
\includegraphics[width=0.23\textwidth]{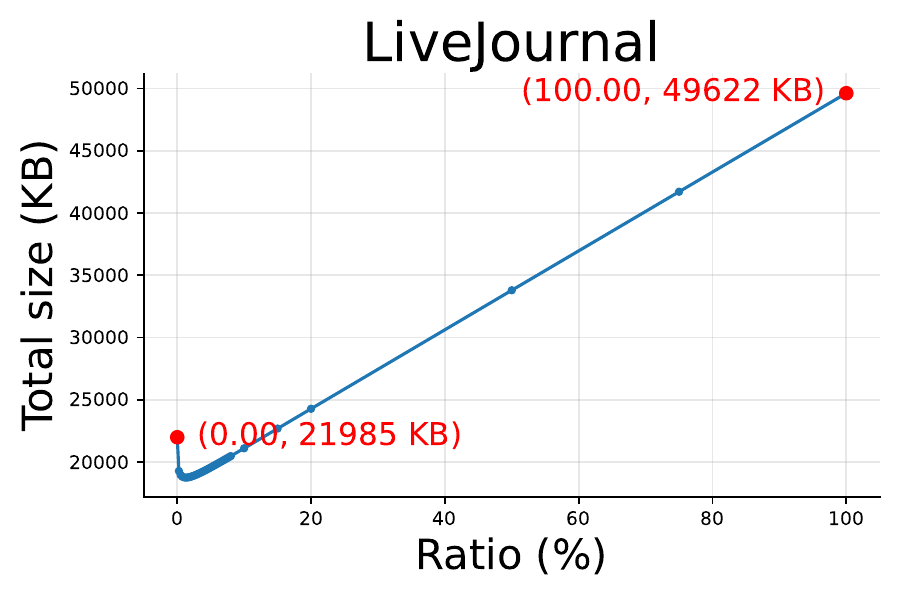}
\includegraphics[width=0.23\textwidth]{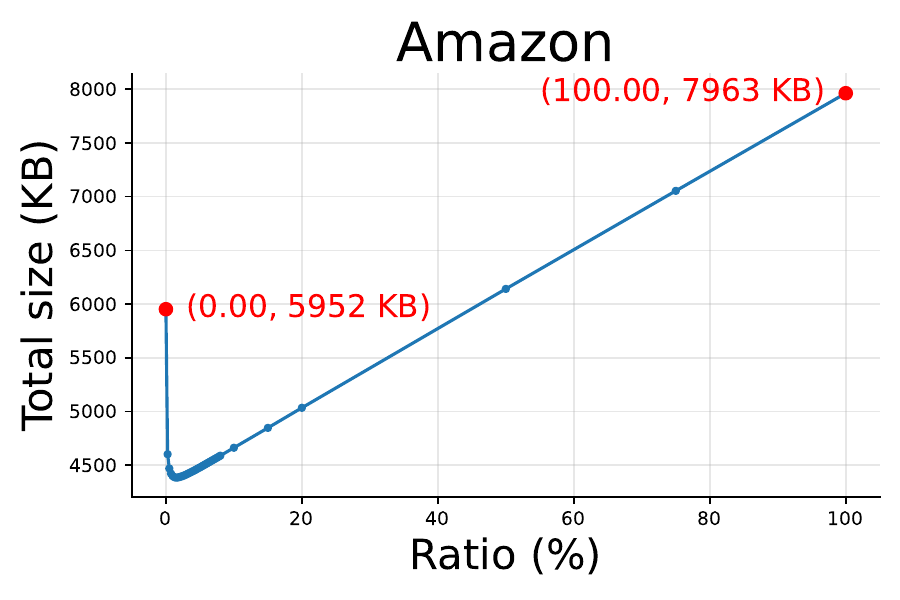}
\includegraphics[width=0.23\textwidth]{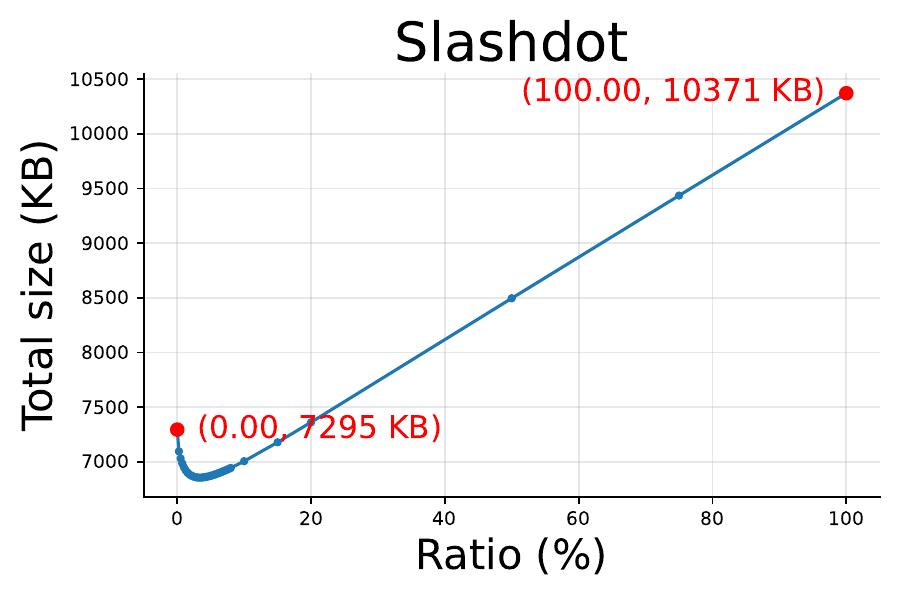}
\caption{Huffman encoding ratio vs. size of compressed hypergraph; (Encoding ratio, Size of compressed hypergraph).}
\label{fig:full}
\end{figure*}

Figure~\ref{fig:full} illustrates the total compressed size of each dataset as the Huffman encoding ratio varies from 0\% to 100\%. Across all four datasets, we observe a consistent U-shaped trend: the compressed size decreases initially, reaches a minimum at a dataset-specific point, and then increases as the Huffman ratio continues to grow. This phenomenon reflects the fundamental trade-off in hybrid encoding between redundancy elimination and prefix overhead.

At $\rho = 0\%$, the encoder defaults entirely to fixed-width bitwise representation. While this scheme incurs no prefix overhead and provides stable performance, it is unable to exploit frequency skew in the symbol distribution. As a result, frequently occurring elements—such as low-degree vertex IDs in power-law graphs—consume more space than necessary.

Conversely, when $\rho = 100\%$, all symbols are encoded using Huffman codes. Although this allows optimal code length for high-frequency items, it significantly penalizes low-frequency values by assigning them long variable-length codes. These long-tail entries not only increase bitstream size but also reduce alignment and decoding efficiency.

Between these extremes lies a dataset-dependent sweet spot. For instance, in Friendster and Amazon, which exhibit highly skewed degree distributions, the optimal point occurs near 10–12\%, indicating that only a small prefix domain needs to be Huffman-encoded to capture most of the entropy gain. In contrast, LiveJournal and Slashdot require a broader Huffman scope due to their flatter frequency profiles. These empirical trends align closely with our theoretical model from Section~\ref{sec:optimal}, where the total encoding cost $C(\rho)$ is shown to have a unique minimizer under mild assumptions.

This U-shaped behavior provides strong evidence for the necessity of hybrid encoding. A fixed strategy—either Huffman-only or bitwise-only—cannot uniformly minimize space across heterogeneous datasets. The ability to dynamically adjust $\rho$ based on symbol statistics is thus essential for achieving robust compression performance.

\subsection{Empirical Validation of Optimality via Regression}
\label{subsec:regression-optimality}

\begin{figure*}[htbp]
\includegraphics[width=0.23\textwidth]{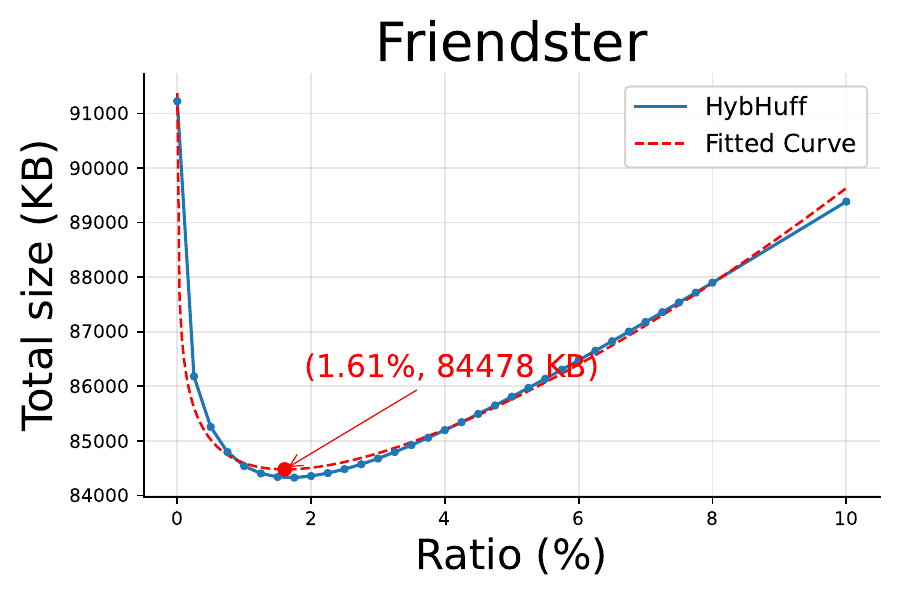}
\includegraphics[width=0.23\textwidth]{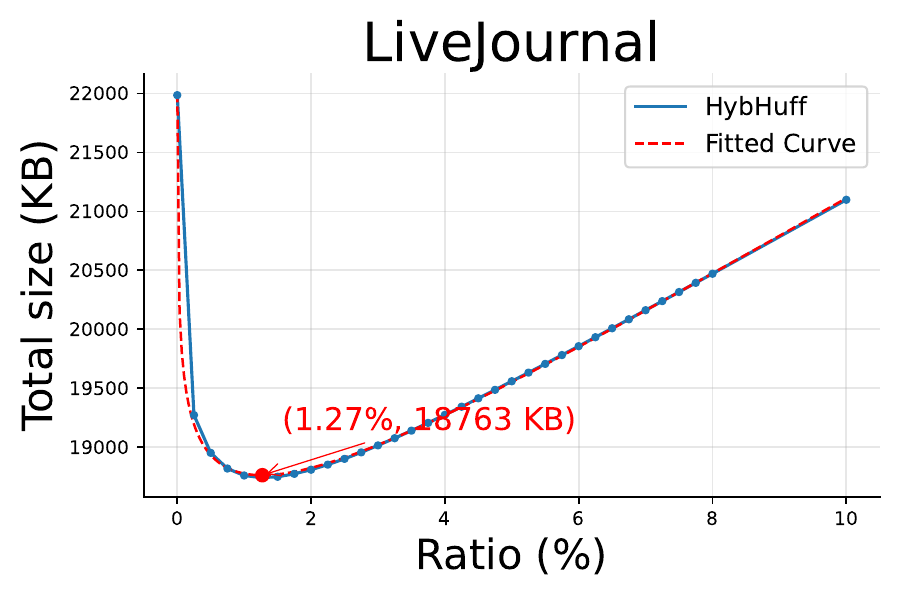}
\includegraphics[width=0.23\textwidth]{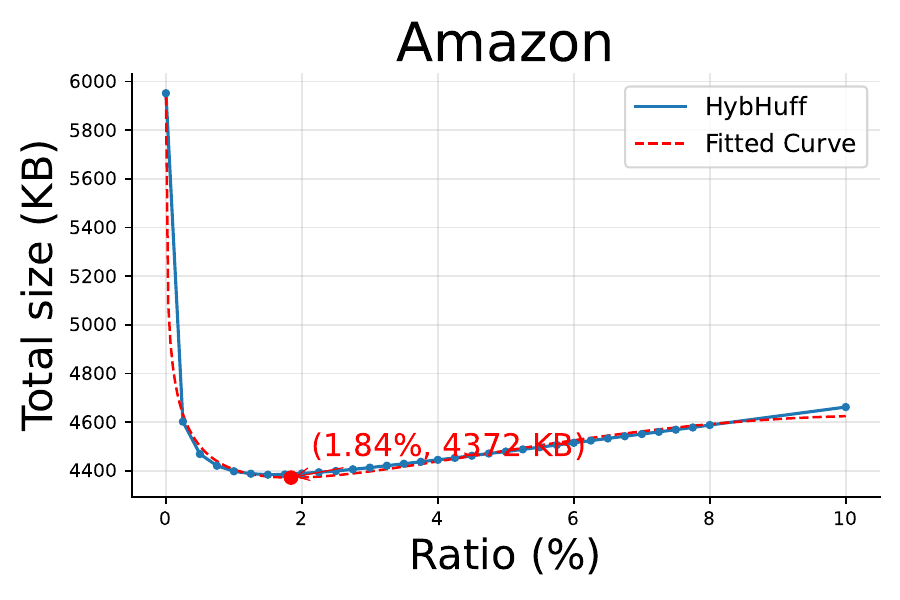}
\includegraphics[width=0.23\textwidth]{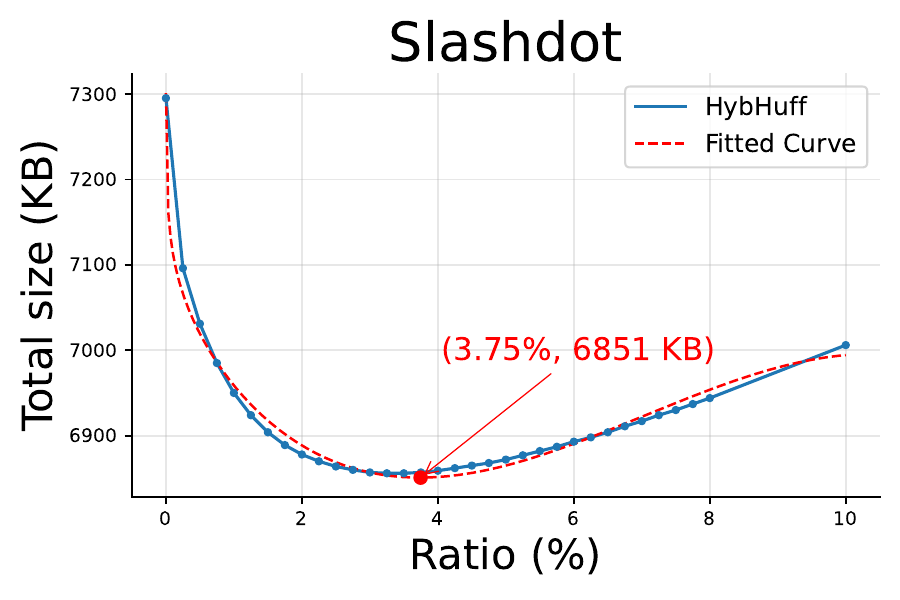}
\caption{Huffman encoding ratio vs. size of compressed hypergraph  (Ratio $\leq 10\%$). The red dots here represent the optimal on the regression curves. (Huffman Encoding Ratio, Size of compressed hypergraph).
}
\label{fig:zoom}
\end{figure*}

To validate the theoretical insights derived in Section~\ref{sec:optimal}, we zoom in on the leftmost 10\% of the Huffman ratio space and perform curve fitting on the observed compressed sizes as shown in Figure~\ref{fig:zoom}. According to Theorem~1 and our lightweight estimation method based on prefix frequency statistics, the total compressed size is expected to follow a non-linear pattern governed by the interaction between entropy reduction and prefix overhead. More concretely, we model the size function using a combination of polynomial and logarithmic components, capturing the initial efficiency gains from encoding high-frequency symbols, followed by a degradation as low-frequency symbols are pulled into the Huffman domain.

We approximate the size function as:
\[
y = a + b x + c x^2 + d \log x + \dots
\]
where $x$ is the Huffman ratio in percentage points. The $\log x$ term dominates in the small-ratio regime, reflecting entropy compression of the most frequent symbols, while higher-order polynomial terms capture the rising overhead as Huffman coverage expands.

Empirically, we observe that this regression model provides an excellent fit to the actual compression behavior across all datasets. For example, the fitted function for Friendster is:
\[
y = 83980.9 + 589.4x + 22.2x^2 - 1069.6 \log x
\]
This curve exhibits a distinct U-shaped profile, with the minimum size attained around $x = 1.61\%$ 
The large negative coefficient on the $\log x$ term explains the sharp initial drop, while the positive $x^2$ term accounts for the upward trend beyond the optimal point. 

The remaining datasets exhibit similar patterns:
\begin{itemize}\small{
  \item \textbf{LiveJournal:} $y = 18359.2 + 420.4x - 2.4x^2 - 527.2 \log x$  
  \item \textbf{Amazon:} $y = 4257.4 + 154.2x - 6.1x^2 - 243.4 \log x$
  \item \textbf{Slashdot:} $y = 7035.5 - 95.1x + 19.4x^2 - 0.9x^3 - 38.4 \log x$
  }
\end{itemize}

Across all four datasets, the negative $\log x$ coefficient is consistent and significant, reinforcing the theoretical claim that early-stage Huffman encoding yields the highest marginal gain per bit. Meanwhile, the positive $x^2$ and $x^3$ coefficients reflect structural overhead as prefix trees grow deeper and symbol distributions flatten.


\subsection{Encoding and Decoding Time vs. Huffman Ratio}
\label{subsec:runtime-vs.-ratio}

\begin{figure*}[htbp]
\includegraphics[width=0.23\textwidth]{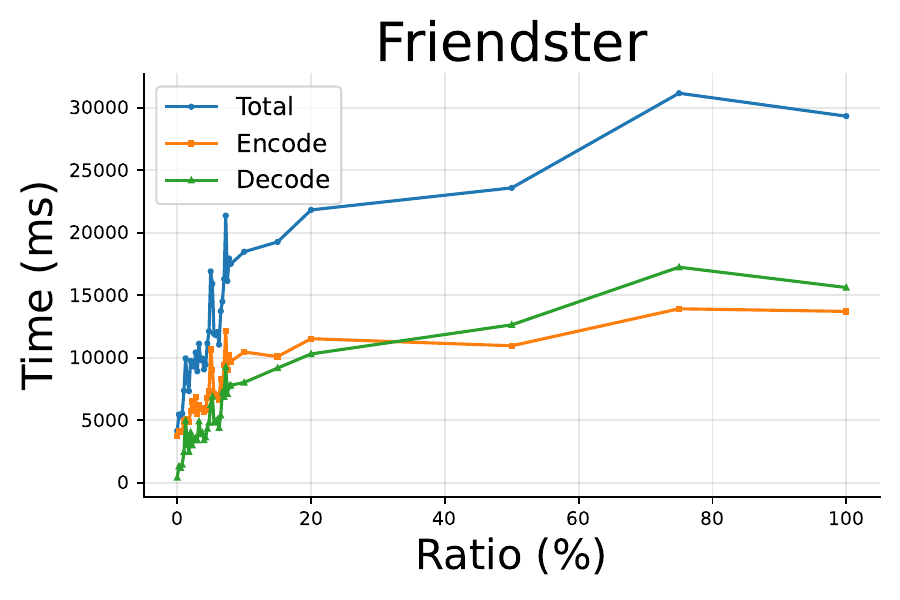}
\includegraphics[width=0.23\textwidth]{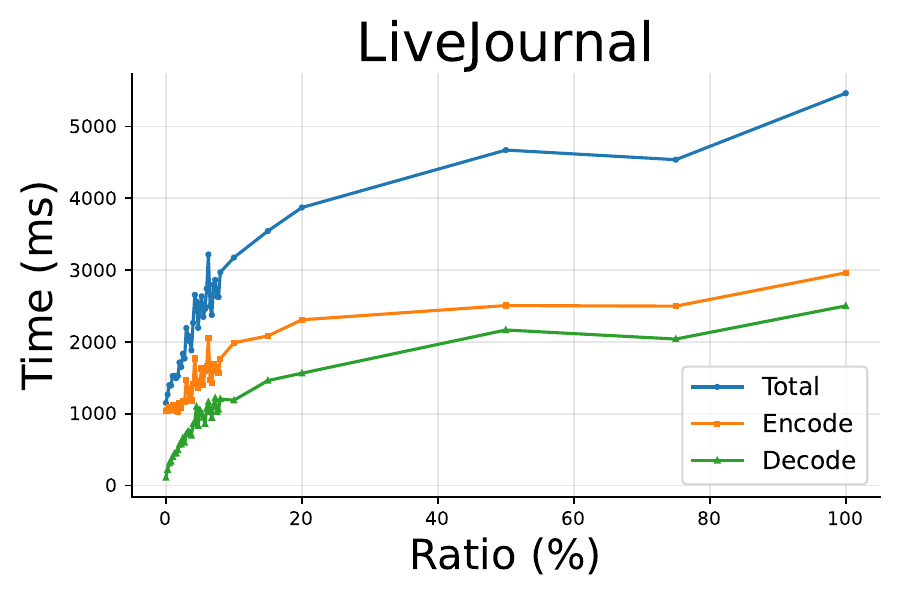}
\includegraphics[width=0.23\textwidth]{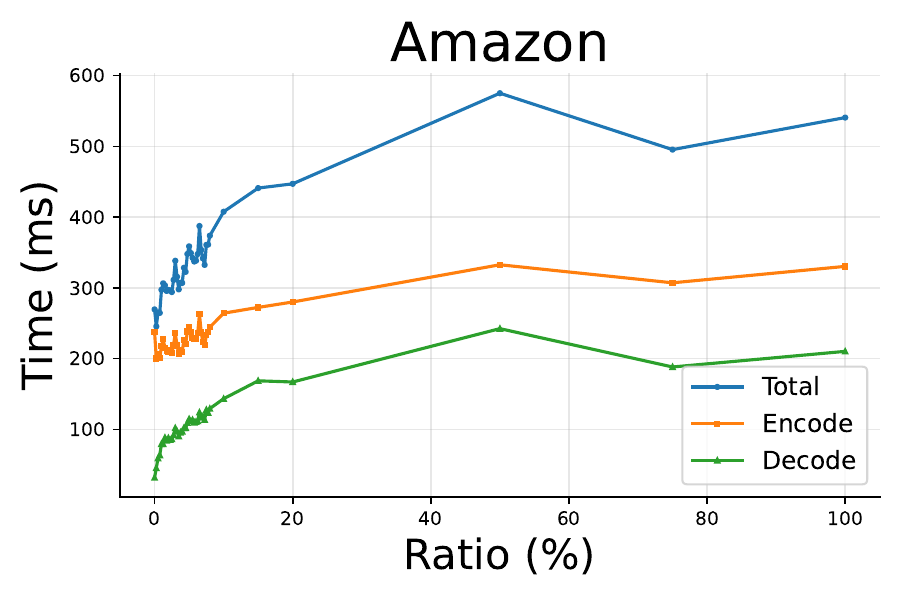}
\includegraphics[width=0.23\textwidth]{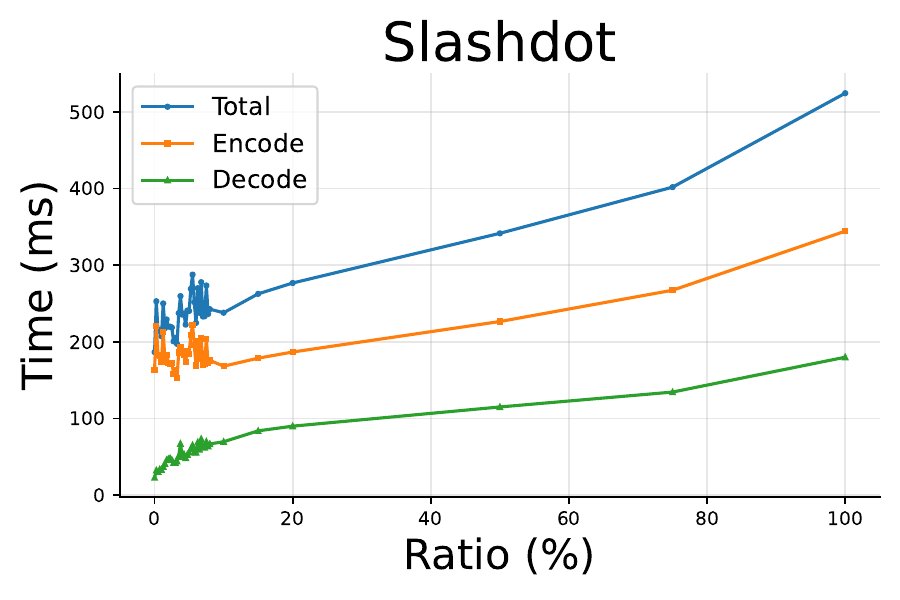}
\caption{Huffman encoding ratio vs. execution time of encoding, decoding, and both.}
\label{fig:time}
\end{figure*}

Figure~\ref{fig:time} shows the execution time of the proposed HybHuff framework as a function of the Huffman encoding ratio. Each subplot reports encoding time, decoding time, and total time across the full range of $\rho$ values. Across all datasets, encoding time increases as more data is assigned to Huffman encoding. This trend is most pronounced in Friendster and LiveJournal, where encoding time roughly triples between $\rho=0\%$ and $\rho=100\%$. The knee around $\rho=5\%$ corresponds to the point where Huffman begins to dominate the stream, causing increased branching, longer codes, and more bit-aligned writes. In contrast, Slashdot and Amazon show smoother growth, likely due to flatter frequency distributions that yield shallower Huffman trees.

Decoding time remains significantly lower and more stable. Even at $\rho=100\%$, all datasets remain under 60 ms. This efficiency comes from our decoding design, which uses precomputed prefix trees, avoids recursion, and decodes bitwise paths with constant-time masking. Minor upward trends, visible in LiveJournal and Slashdot at high $\rho$, stem from deeper Huffman trees introducing longer traversal paths for low-frequency symbols.

The widening gap between encoding and decoding time across the $\rho$ axis highlights an important asymmetry: the cost of Huffman encoding scales rapidly due to prefix lookup and alignment logic, while decoding is bounded by tree depth and optimized scan paths. Amazon, being the smallest dataset, demonstrates that absolute overheads remain manageable even at high ratios.

From a system design perspective, these results emphasize the tunability of our hybrid scheme. By choosing $\rho$ near the compression-optimal point (see Figure~\ref{fig:compression_rate}), one can achieve compact output while keeping both encoding and decoding within acceptable latency budgets. For use cases where encoding speed is critical, a conservative $\rho$ (e.g., $<5\%$) yields fast, low-overhead operation with modest compression loss.

\subsection{Application-Level Runtime Analysis}
\label{subsec:application-runtime}

\begin{figure*}[htbp]
\includegraphics[width=0.23\textwidth]{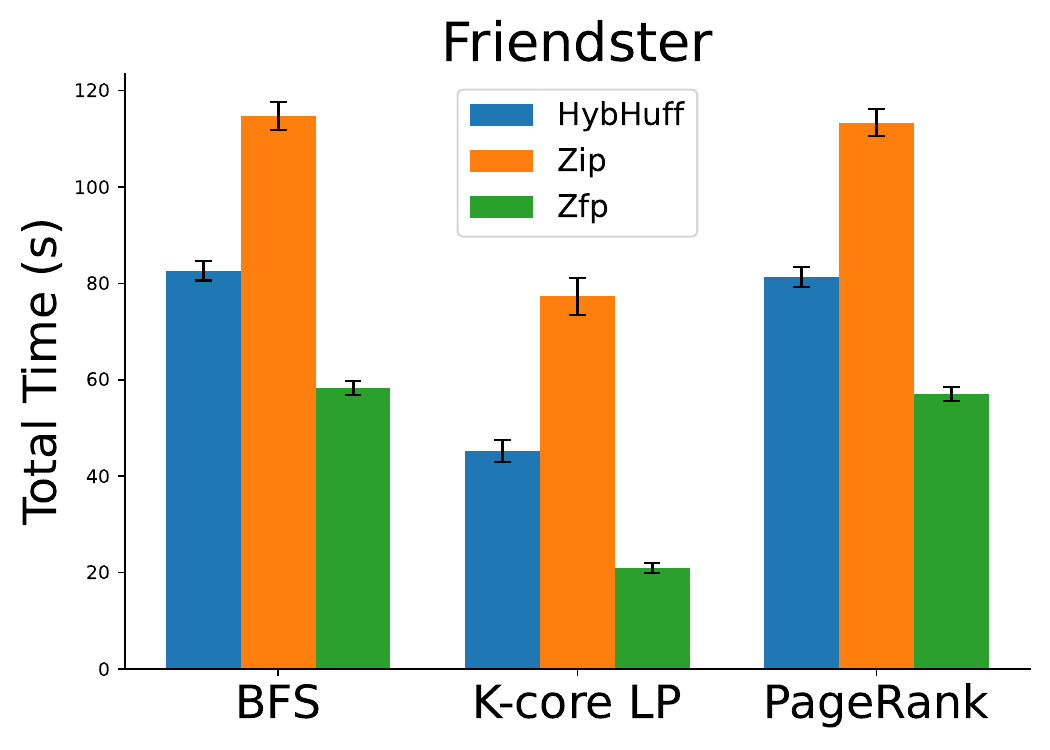}
\includegraphics[width=0.23\textwidth]{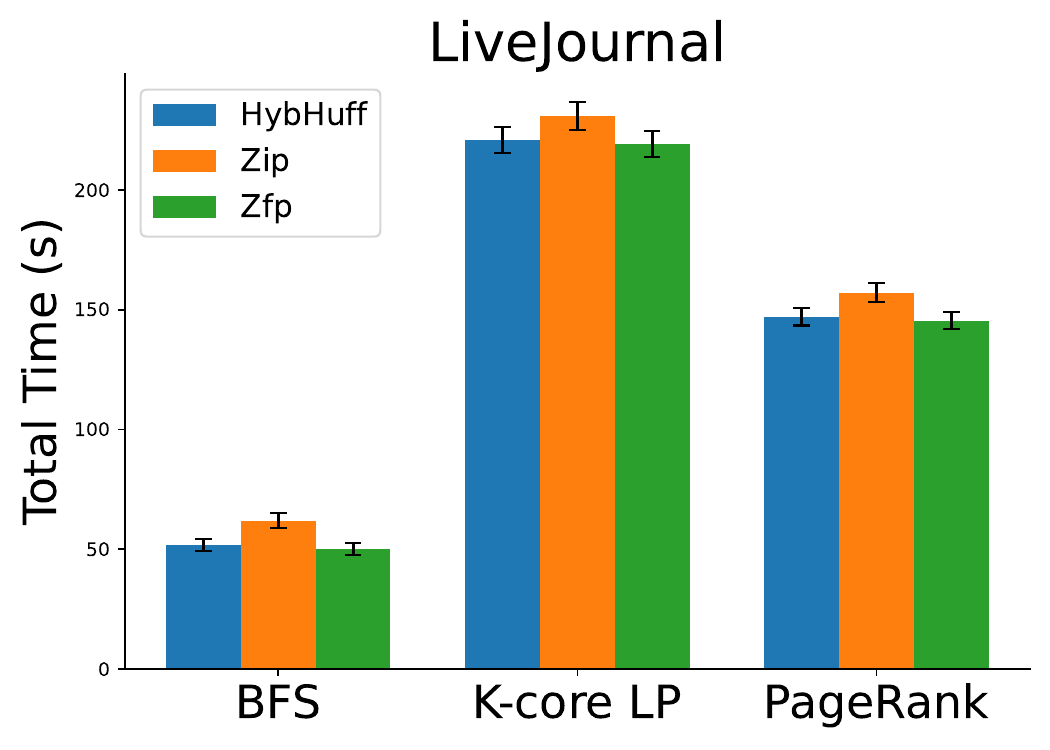}
\includegraphics[width=0.23\textwidth]{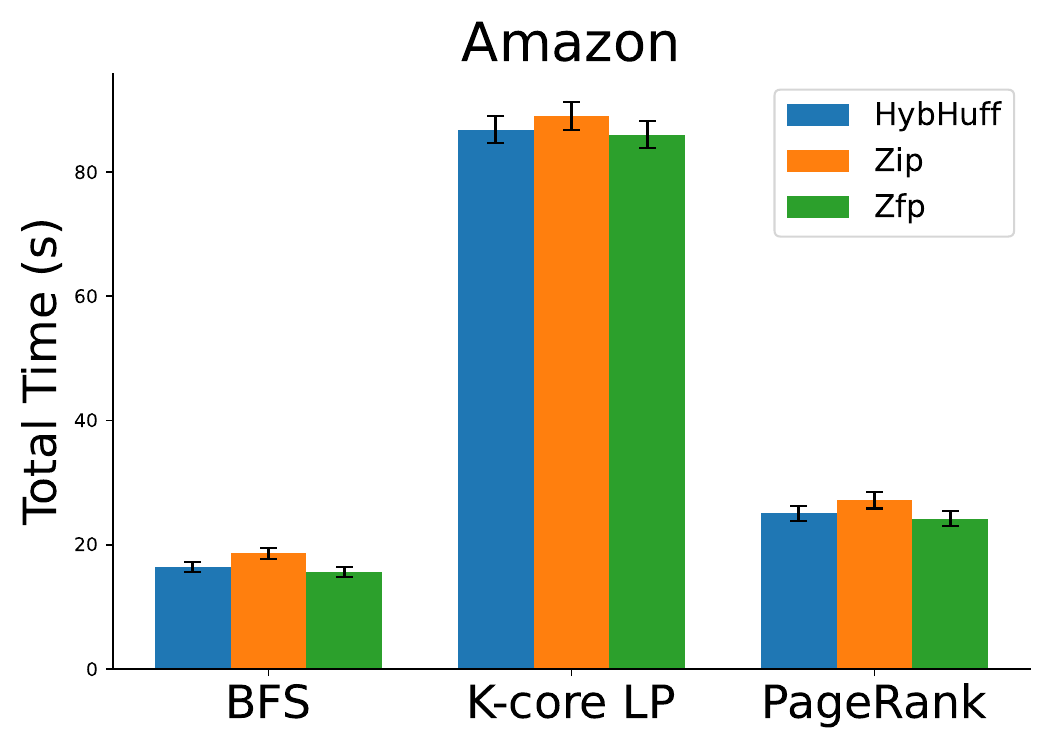}
\includegraphics[width=0.23\textwidth]{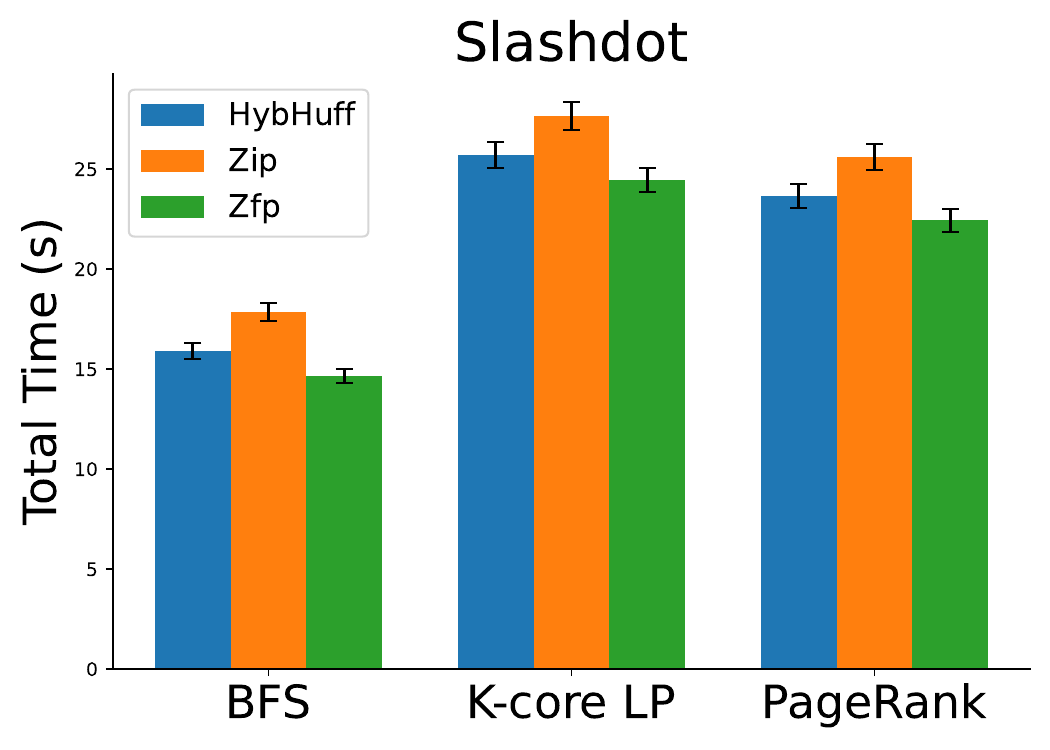}
\caption{Comparison of time with compression methods across hypergraph applications, including both the decompression time and the application execution time. 
}
\label{fig:ApplicationTime}
\end{figure*}

We now assess the practical utility of \textit{HybHuff} by measuring end-to-end runtime in three representative hypergraph applications: BFS, PageRank, and $k$-core label propagation. Each workload operates directly on the compressed representation via an iterator-based interface. Figure~\ref{fig:ApplicationTime} shows the total runtime of these applications under three compression schemes across four datasets.

In most cases, HybHuff matches or outperforms \texttt{zip} and \texttt{ZFP}. For example, in LiveJournal and Slashdot, HybHuff consistently yields the fastest or second-fastest runtimes across all tasks. In Amazon, all methods perform similarly due to the dataset’s small size. The only exception occurs in the BFS task on Friendster, where \texttt{ZFP} shows slightly better performance. This is likely due to ZFP’s low per-block decoding overhead and the high spatial regularity of Friendster’s 2-section structure, which aligns well with ZFP’s floating-point block compression model.

It is important to emphasize, however, that runtime is not the primary objective of our framework. HybHuff is designed to achieve high compression while maintaining access efficiency. As shown earlier in Figure~\ref{fig:compression_rate}, HybHuff achieves up to $2.3\times$ smaller compressed size than \texttt{zip} and $1.9\times$ smaller than \texttt{ZFP}, while still maintaining competitive or superior runtime in most scenarios.

In BFS and $k$-core, our streaming decode model supports low-latency adjacency access, even under irregular traversal patterns. PageRank, which is sensitive to layout and memory reuse, benefits particularly from our metadata-guided access model. Across all tasks, the amortized decoding cost remains in the microsecond range.


\subsection{Discussion}

\paragraph{Encoding Ratio vs. Runtime Tradeoff} 
As shown in Figures~\ref{fig:compression_rate} and~\ref{fig:ApplicationTime}, HybHuff consistently achieves higher compression ratios than both \texttt{zip} and \texttt{zfp} across all datasets. Although \texttt{zfp} can occasionally offer faster encoding or application-level runtime (e.g., BFS on Friendster), these gains are dataset-specific and stem from block-wise optimizations. Overall, HybHuff’s entropy-guided coordination balances space savings with competitive runtime, offering a more robust and generalizable tradeoff.

\paragraph{Threshold Sensitivity and System Integration}
Figures~\ref{fig:full} and~\ref{fig:zoom} highlight the existence of a dataset-dependent optimal Huffman ratio, reflected by a consistent U-shaped compression trend. This optimal point can be reliably estimated via lightweight regression, opening the door to automated tuning. HybHuff also exposes a stateless decoder interface supporting eager and lazy evaluation, facilitating integration with hypergraph systems under various resource constraints, including streaming and pipelined execution.

\section{Conclusion}

This work presents a hybrid compression framework for hypergraph adjacency that synthesizes Huffman and bitwise encoding through entropy-aware coordination. By leveraging the inherent skew and sparsity in real-world hypergraphs, our method adaptively partitions the symbol space to minimize total space cost. We formalize this coordination as an optimization problem, prove the existence of an optimal threshold, and develop a practical estimation strategy grounded in frequency statistics. Extensive experiments demonstrate that our approach consistently outperforms general-purpose and domain-specific compressors in compression ratio, while preserving fast decoding and seamless integration with popular workloads. 
Beyond improving memory efficiency, this work opens new possibilities for scalable hypergraph analytics under resource constraints, bridging the gap between structural compactness and algorithmic performance. We hope it serves as a step toward principled, system-level support for higher-order relational modeling in large-scale data systems. 

\section*{Acknowledgments}
This research is supported by the U.S.\@ Department of Energy (DOE) through
  the Office of Advanced Scientific Computing Research's ``Orchestration for Distributed \& Data-Intensive Scientific Exploration'' and
  the ``Cloud, HPC, and Edge for Science and Security'' LDRD at Pacific Northwest National Laboratory.
PNNL 
is operated by Battelle for the DOE under Contract DE-AC05-76RL01830.
Results presented in this paper were partly obtained using the Chameleon testbed supported by the National Science Foundation.






%


\bibliographystyle{abbrv}
\bibliography{main}



\end{document}